\documentclass[sigconf, nonacm]{acmart}

\AtBeginDocument{%
  \providecommand\BibTeX{{%
    \normalfont B\kern-0.5em{\scshape i\kern-0.25em b}\kern-0.8em\TeX}}}

\usepackage{color}
\usepackage{xcolor}
\usepackage{qtree}
\usepackage{graphicx}
\usepackage{algorithm}
\usepackage[noend]{algpseudocode}
\usepackage{subcaption}
\usepackage{stmaryrd}
\usepackage{amsmath}
\usepackage{array}
\usepackage{amsfonts}

\usepackage{xspace}
\usepackage{setspace}
\usepackage{enumitem}
\newcommand{\braces}[1]{\left \lbrace {#1} \right \rbrace}
\usepackage{pgfplots}
\usepackage{filecontents}
\pgfplotsset{compat=1.14}

\definecolor{codegreen}{rgb}{0,0.6,0}
\definecolor{ao}{rgb}{0.0, 0.5, 0.0}
\definecolor{codegray}{rgb}{0.5,0.5,0.5}
\definecolor{codepurple}{rgb}{0.58,0,0.82}
\definecolor{backcolour}{rgb}{0.95,0.95,0.92}

\newcommand{\nb}{\textsc{NBLyzer}\xspace}

\newtheorem{obs}{Observation}[section]

\algdef{SE}[VARIABLES]{Variables}{EndVariables}
   {\algorithmicvariables}
   {\algorithmicend\ \algorithmicvariables}
\algnewcommand{\algorithmicvariables}{\textbf{global variables}}

\def\cccolorbox#1#2{\ifx#2\relax\let\next\allowbreak\else
       \def\next{\colorbox{#1}{#2}\allowbreak\cccolorbox{#1}}\fi\next}
\def\ccolorbox#1#2{\fboxsep0pt\cccolorbox{#1}#2\relax}

\def\!#1{\ifx#1\ccolorbox\allowbreak\expandafter\ccolorbox\else
         \ifx#1\end\expandafter\expandafter\expandafter\end\else
         #1\allowbreak\expandafter\expandafter\expandafter\!\fi\fi}

\def\cccolorbox#1#2{\ifx#2\relax\let\next\allowbreak\else
   \def\next{\colorbox{#1}{\strut #2}\allowbreak\cccolorbox{#1}}\fi\next}

\begin{document}

\title{A Static Analysis Framework for Data Science Notebooks}

\author{Pavle Suboti\'{c}}
\email{pavlesubotic@microsoft.com}
\affiliation{%
  \institution{Microsoft}
  \country{}
}

\author{Lazar Miliki\'{c}}
\email{al-milikic@microsoft.com}
\affiliation{%
  \institution{Microsoft}
  \country{}
}


%

\author{Milan Stoji\'{c}}
\email{milan.stojic@microsoft.com}

\affiliation{%
  \institution{Microsoft}
  \country{}
}


\begin{abstract}

Notebooks provide an interactive environment for programmers to develop code, analyse data and inject 
interleaved visualizations in a single 
environment. Despite their flexibility, a major pitfall that data scientists encounter is unexpected behaviour 
caused by the unique out-of-order execution model of notebooks. As a result, data scientists face various 
challenges ranging from notebook correctness, reproducibility and cleaning. In this paper, we propose a 
framework that performs static analysis on notebooks, incorporating their unique execution semantics. Our 
framework is general in the sense that it accommodate for a wide range of analyses, 
useful for various notebook use cases. We have instantiated our framework on a diverse set of analyses and 
have evaluated them on 2211 real world notebooks. Our evaluation demonstrates that the vast majority (98.7\%) 
of notebooks can be analysed in less than a second, well within the time frame required by interactive notebook clients.
\end{abstract}






\maketitle

\section{Introduction}
Notebooks have become an increasingly popular development environment for data science. In 2018, GitHub reported 2.5 million 
notebooks, an increase from 200K in 2015~\cite{notebookscience}. Notebooks provide 
a dynamic read-eval-print-loop (REPL) experience where 
developers can rapidly prototype code while interleaving data visualizations including graphs, textual 
descriptions, tables etc.  A 
notable peculiarity of notebooks is that the program i.e., notebook, is divided into non-scope inducing blocks of 
code called \emph{cells}. Cells can be added, 
edited and deleted on demand by the user. Most importantly, cells, regardless of their order in the notebook,
can be executed (and re-executed) by the user in \emph{any given sequence}. This provides a level of incrementalism that 
improves productivity and flexibility. At the same time, such execution semantics 
make notebook behaviour notoriously difficult to 
predict and reproduce. This observation is highlight in~\cite{zeller2}. Here the difficulty of reproducing notebook results 
is investigated, concluding that from a large set of notebooks, only 25\% of notebooks could be executed without an 
error and less than 5\% were trivially reproducible. Moreover, in~\cite{zeller} it was observed that there is 
an abundance of code smells and bugs in 
real world notebooks. Thus the authors of this study argue for more code analysis tooling to improve the quality of notebooks. To understand the source of this phenomena consider the example below.

\begin{figure*}
    \centering
    \begin{subfigure}[b]{0.33\linewidth}
    \includegraphics[width=\textwidth]{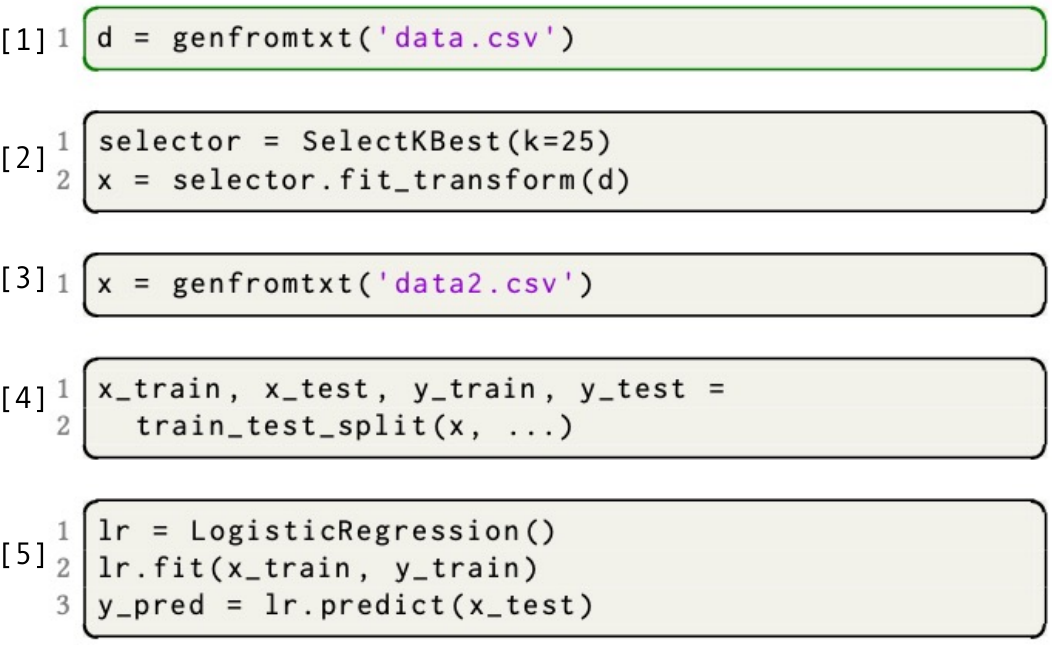}
    \caption{Initial notebook\label{fig:ex1}}
    \end{subfigure}\hfill
    \begin{subfigure}[b]{0.33\linewidth}
    \includegraphics[width=\textwidth]{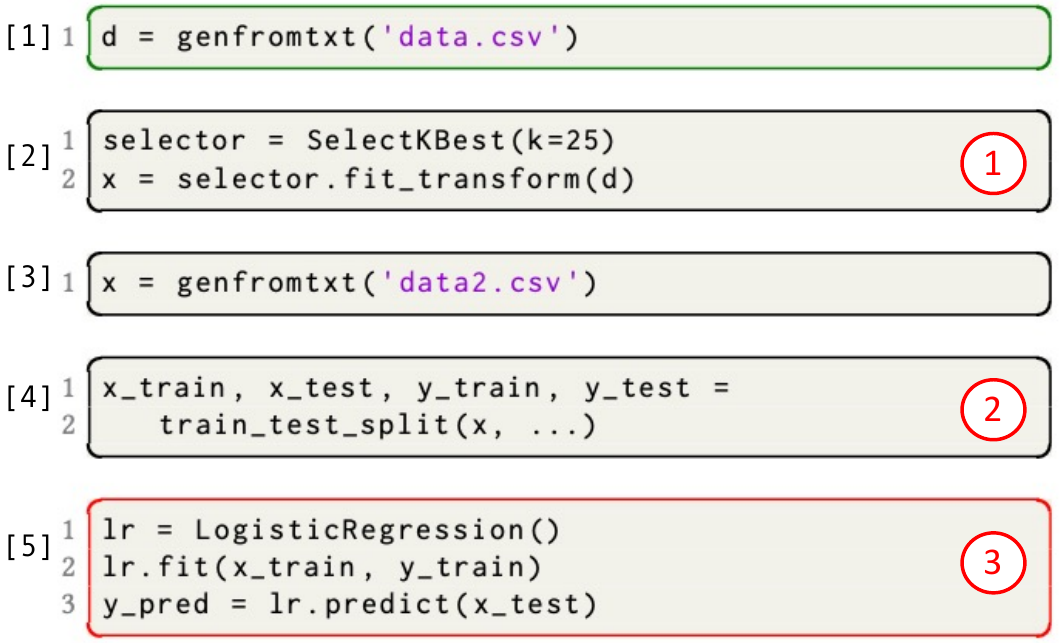}
    \caption{Data leak analysis\label{fig:ex2}}
    \end{subfigure}\hfill
    \begin{subfigure}[b]{0.33\linewidth}
    \includegraphics[width=\textwidth]{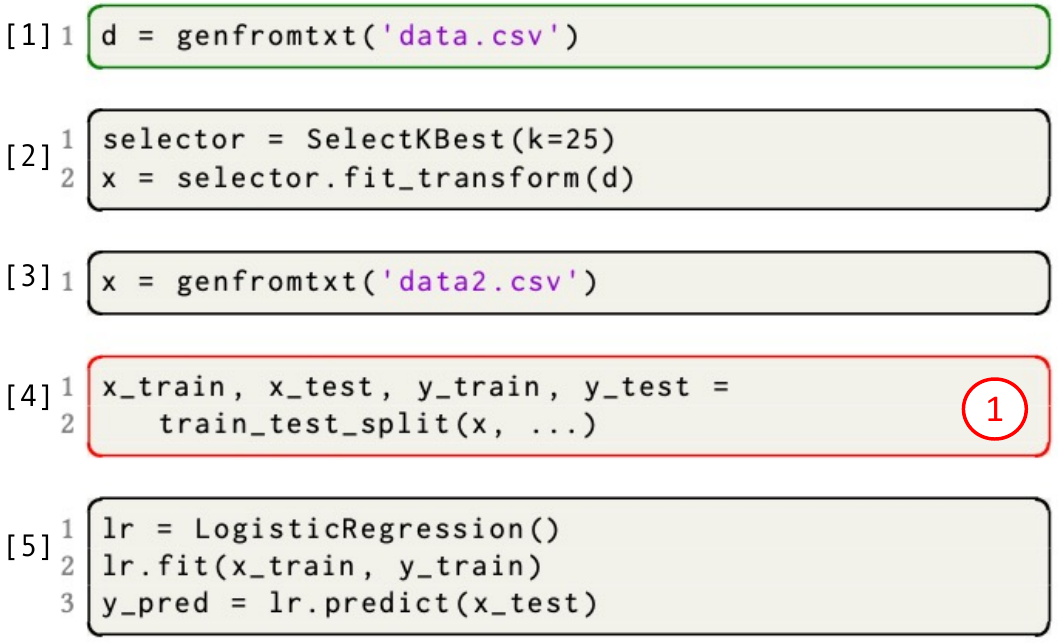}
    \caption{Stale state analysis\label{fig:ex3} }
    \end{subfigure}
    \caption{Example Notebook\label{fig:motivating}}
\end{figure*}

\begin{example}[Motivating Example]
\label{ex:motivating}
Consider the notebook in Figure~\ref{fig:motivating} comprising of $5$ cells.
Each numbered from $1$ to $5$, indicated by the left hand side number in square brackets. In a script, the 
execution proceeds as if the cells were merged into 
a single cell and each statement is executed as dictated by the regular program control flow i.e., 
statements in \textit{cell 1} are executed sequentially, followed by \textit{cell 2}, \textit{cell 3} and 
so on. However, in notebooks any given cell can be executed at any given time by the user. This produces 
a potentially infinite space of possible execution paths 
due to a lack of constraints on the order in which cells can be executed in. 
In the example notebook, \textit{cell 1} and \textit{cell 3} read data from 
a file into a data frame. In \textit{cell 2} data in variable \texttt{d} is normalized
and in \textit{cell 4} the data is split into test and training segments. In \textit{cell 5} 
the model is trained, tested and assessed for accuracy. It is apparent that several legitimate executions 
exist in this notebook. For example, one could start by executing \textit{cell 3}, \textit{cell 4} and 
\textit{cell 5}. Another execution sequence is \textit{cell 1}, followed by \textit{cell 2}, \textit{cell 4} and 
\textit{cell 5}. Note, it is common to have several alternate executions in a notebook for reasons of 
experimentation etc.

Now consider the following scenario: suppose the user executes the sequence of 
cells $1$, $2$, $4$, and $5$. While this may not seem particularly ominous, it will in fact result in a 
data leak bug~\cite{dataleak} as the function in 
\textit{cell 2} normalizes the data and yet \textit{cell 4} splits the data in to train and test
data \emph{after} the normalization, thus resulting in a what is know in data science as a data leak. Now, lets suppose the user, 
after some investigation, identifies this problem. They re-execute \text{cell 1}, skipping \text{cell 2} (to avoid 
normalization) and execute \textit{cell 4} and \textit{cell 5}. The user may be perplexed as the same issue 
re-occurs. The problem is that 
the user executed \textit{cell 4} which referred to variable $x$, which was previously computed by 
\textit{cell 2}, resulting the using an old (or stale) value. One can see how a user can quickly get into a 
confusing situation, even for this relatively simple notebook\footnote{in fact this example is based on a real 
life notebook sourced from stackoverflow, where a user was experiencing a data leak}.

\end{example}

Each of the bugs above demonstrate the ease at which a seemingly simple data science program can 
result in unforeseen behaviour in a notebook environment. Moreover, establishing the root cause is 
similarity difficult without engaging in time consuming post-mortem debugging that cannot prevent 
future bugs, or generalize the cause of the bug. On the other hand, if we were to restrict
the notebook execution semantics, we would be removing precisely the flexibility that have made notebooks 
so popular in the last few years. 

In this paper, we argue for the use of static analysis as an ante-mortem debugging mechanism. Our technique, 
co-exists with the unique notebook execution flexibility, and yet reducing errors 
and debugging efforts by warning users \emph{ahead of time} of hypothetical erroneous and 
safe actions. Due to the semantics of notebooks, standard static analysis tools 
cannot be directly applied, and to the best of our knowledge general static analyzers targeting notebooks have not been proposed 
in the literature. Moreover, the interactive nature of notebooks requires analyzers that are able to 
produce feedback within a second~\cite{rail} to not disrupt the flow of the user experience. To this end, we 
propose \nb, a static analysis framework that 
provides notebook users the ability to perform a
\emph{what-if} static analysis on given notebook actions. Notebook actions constitute opening a notebook, 
code changes, cell executions, cell creation, deletion etc. Our framework soundly reports potential 
issues that may occur for a given action within the time bounds required for users not to notice a  
disruption. For instance, consider the notebook in Example~\ref{ex:motivating},
as shown in Figure~\ref{fig:ex2}, our framework will warn the user that the event of executing \emph{cell 1} can 
lead to a data leakage by executing cells \textit{2, 4} and \textit{5} in a milliseconds.  Moreover, as shown in 
Figure~\ref{fig:ex3},  it will warn in milliseconds that the event of executing \textit{cell 1}
can result in a stale state if \textit{cell 4} is executed before \textit{cell 2}. Conversely, our framework can  
recommend the execution of the sequence\textit{cell 3, 4, 5} to be safe to execute after \textit{cell 1} is executed. Our 
framework supports a wide range of static analyses\footnote{that can be expressed as an abstract interpretation}. In 
Section~\ref{sec:analyses} we present several analyses 
implemented in \nb, particularly targeting data science notebooks. Using these analyses, we 
facilitate the automation of several other important notebook development use cases 
including, notebook correctness, notebook debugging, notebook reproducibility, notebook auditing, and notebook cleaning.

\nb employs the theory of Abstract 
Interpretation~\cite{CC77} to perform intra-cell static analyses i.e., on individual cells, and thus 
in-cell termination is guaranteed 
for the price of an over-approximate analysis result. The key idea is to over-approximate the notebook semantics and computational state $\sigma$ 
and instead produce an abstract state $\sigma^\sharp$ which comprises of an element of an \emph{abstract domain} that 
encodes the analysis property of interest.  When analyses are triggered by an event, 
an inter-cell analysis is performed by propagating the analyses results (abstract state) to valid successor cells 
in the notebook. To select valid successors, we introduce the notion of 
\emph{cell propagation dependencies}, which prunes away unnecessary sequences of 
cell executions on-the-fly, parametrized by the current abstract state. In this way, the abstract state is propagated 
efficiently while ensuring soundness and termination via both an intra and inter-cell termination criteria. We evaluate 
\nb on 2211 real world notebooks and several instantiated analyses to demonstrate its utility and scalability. \nb is able to 
analyze 98.7\% of notebooks within a second. Overall, 
we claim the following contributions: 

\begin{enumerate}
\item A novel static analysis framework specifically for notebooks
\item Several instantiated industrial static analyses implemented in \nb, including a novel data leakage analysis that 
to the best of our knowledge, has not been presented before in the literature or implemented in any static analyzer.
\item An extensive performance evaluation of \nb that demonstrates its adequate performance based on~\cite{rail} 
to perform only-the-fly static analysis within the notebook environment without degrading the user notebook experience.
\end{enumerate}

\section{Overview}
In this section we give an overview of the \nb static analysis framework for notebooks. 
\begin{figure}
    \centering
    \includegraphics[width=0.35\textwidth]{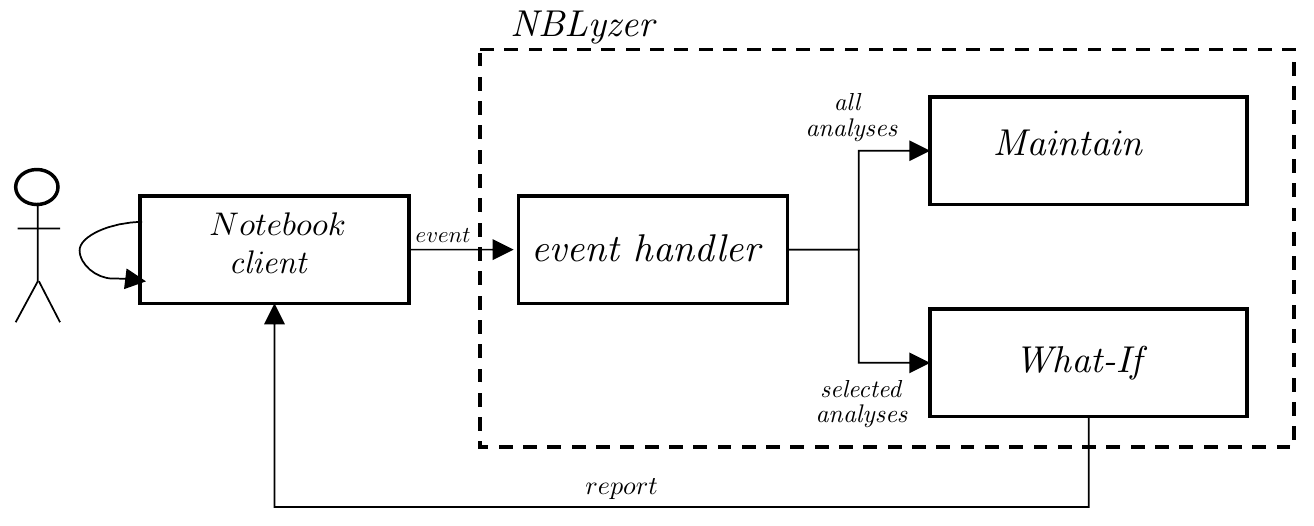}
    \caption{Overview of \nb \label{fig:overview}}
\end{figure}

The high-level operation of the framework is depicted in Figure~\ref{fig:overview}. A user performs actions on 
a notebook such as opening notebooks, adding cells, changing cells, executing cells, deleting cells 
among others. For each event, the user may want to initiate a what-if analysis, essentially asking \nb 
\emph{what can occur if I execute cell 1?}. This allows the user to ahead of time avoid putting the 
notebook in a state that will lead to an error. Conversely, the user may ask \nb 
\emph{What executions are safe if I execute cell 1?}. This allows the user to choose from a set of possible 
safe execution paths. Other examples of what-if questions include 
\emph{Which cells will become isolated if I rename d to x cell 2?}, \emph{Which cells will are redundant in the notebook I opened?}
etc.

Each of these what-if questions can be useful for use cases including reproducibility, security auditing, 
notebook cleaning and simplification, debugging and education among others.

From a systems perspective, a what-if analysis is a notebook event that is associated/configured to 
a set of analyses. For example, asking about notebook safety will entail a certain set of analyses, 
and asking about notebook cleanliness will entail a different set of analyses. \nb therefore intercepts 
an event from the notebook client and determines the appropriate mode of operation. The modes of operation are 
described below.

\paragraph{\textbf{Maintenance mode.}}
In the case that the event is an cell execution and the user has not no attached analyses to this event, i.e., a what-if analysis is not triggered, 
then \nb will perform cell maintenance for the executed cell. Since a cell execution will result in the concrete notebook 
state being updated, \nb needs to ensure that the corresponding abstract state for all future invoked analyses is maintained. In addition 
code summaries that enable faster analyses must also be updated.
\nb performs maintenance by updating (if the code has changes) all intermediate program representations including parsing 
the cell code into an abstract syntax tree (AST), converted the AST to a 
control flow graph (CFG) and producing use definition (U-D) chains. If the cell code has not changed, 
these are retrieved from a cache. Using the CFG the static analyses are performed to update the notebook's abstract state, i.e., the resultant state 
from a static analysis required to perform future static analyses in the future. In the case the event is a non-execution event an abstract state may 
not need to be computed and only summary, AST, CFG and U-D information is updated. In Section~\ref{ss:intra-technical} we provide a 
detailed account of the maintenance process. 

\paragraph{\textbf{Propagation mode.}} In the case of a what-if analysis, i.e., an event with 
a subset of analyses associated with it, a inter-cell analysis is performed. Here starting from 
the global notebook abstract state, a set of possible abstract states are computed corresponding to 
the set of possible executions up to a limit $K$ depth or a fixpoint 
is reached on all branches. This process 
for each cell, checks which other cells have a dependency and propagate the computed abstract state 
to the dependent cells, for which the incoming abstract state is treated as a initial state. For each cell 
the abstract state is checked for correctness criteria, if a error is found a report is updated which 
serves as instruction for notebook clients to alert the user to the consequences of the event (e.g., by cell highlighting etc.). In 
Section~\ref{ss:inter-technical} we provide a detailed account of the maintenance process.




\section{Technical Description}
\label{sec:technical}
In this section we provide a technical description of the \nb framework.

\subsection{Notebook Program Model}

\subsubsection{Notebook}
A Notebook $N$ consists of a set of cells 
$c_i \in N$.  A cell $c_i$ consists of a sequence of 
code statements $st^i_j(l, l')$ from a location $l$ to location $l'$ in a 
control flow graph (CFG) where $i$ refers to the cell number and $j$ the index of 
the statement in the cell. As an abuse of notation we allow $c_i$ to be also 
used as a label.

\subsubsection{Cell Execution}
An execution of a cell $c_i$ over a state space $\Sigma = V \rightarrow D$ where $V$ is the 
set of notebook variables and $D$ is the concrete domain of execution, is denoted by 
$\sigma_{i+1} = \llbracket c_{i+1} \rrbracket (\sigma_{i})$, assuming the execution of $c_{i+1}$ terminates. Here $\sigma_{i+1} \in \Sigma$ is the 
output state and $\sigma_{i} \in \Sigma$ is the input state previously computed by a cell 
$c_i$ in the \emph{execution sequence}. We denote access into a state as $\sigma(v)$ where $v \in V$ and we denote updating a state 
as $\sigma[v \mapsto d]$ where $v \in V$ and $d \in D$.

\subsubsection{Notebook Execution}
A notebook execution is a potentially infinite execution sequence 
$\sigma_0 \rightarrow_{c_i} \sigma_{1} \rightarrow_{c_j} \dots$ where $\forall k \geq 0, c_k \in N, \sigma_k \in \Sigma$ 
and $i = j \vee i \neq j$. The choice of the next cell in a execution sequence is determined by the user from the space of 
all cells in a notebook.

\subsection{Analysis Framework}

\subsubsection{Intra Cell Analysis}
\label{ss:intra-technical}
\paragraph{Events and Analyses.}
The inter-cell analysis is triggered by an event $e \in Event$. An event is attached to a set of 
analyses $A' \subset A$ by an mapping $\mathcal{M}: \textit{Event} \rightarrow \wp(A)$.
An analysis $a$ is a tuple of a abstraction label $abs$ and condition $cond$. The condition 
$cond$ is an assertion on an abstract state of the analysis of type $abs$. 

\paragraph{Abstract state computation.} From the sequence of statements in a cell, 
we construct a control flow graph (CFG), a directed graph that encodes the 
control flow of the statements in a cell. We define a CFG as $\langle L, E \rangle$ where 
an edge $(l, st, l') \in E$ reflects the semantics of the cell statement $st$ associated with the CFG edge from 
locations $l$ to  $l'$ in the cell.

A sound over-approximation $\sigma^\sharp$ of a state $\sigma$ is computed by 
iteratively solving the semantic fixed point equation
$\sigma^\sharp = \sigma^{\sharp}_0 \sqcup \llbracket \bar{st} \rrbracket^\sharp(\sigma^\sharp)$ using the abstract semantics
$\llbracket \bar{st} \rrbracket^\sharp$ for statements $\bar{st}$ in a cell and the initial abstract 
state ($\sigma^\sharp_0$). At the cell level this 
computation is defined as $F_{c_i}$ which we refer to a \emph{abstract cell transformer}. $F_{c_i}$ takes an abstract state
and computes a fix-point solution~\cite{fix, max} in the abstract domain.

Since a what-if analysis may not be triggered on every event, and yet a cell is executed by the user, it is of small
cost to maintain the abstract state along with the concrete state, as our analyses are design to be 
faster than performing a concrete execution (See Section~\ref{sec:evaluation}). We therefore maintain an abstract state $\sigma^\sharp$ 
which is updated each time a cell is executed, in parallel to 
the concrete executions of a notebook cell. At each 
execution, a cell transformer $F_{c_i}$ for a cell $c_i$ is applied with the current global state, returning a updated 
new global state i.e., $F_{c_i}(\sigma^\sharp) =  \sigma^{\sharp\prime}$.  We perform this maintenance for two reasons: Firstly, we may want to perform a static analysis 
just before cell execution, blocking execution if an error was found. Secondly, the global abstract state is needed 
to initiate a what-if analyses, once its triggered by a user.


To analyze a cell, we reduce the static analysis problem to the computation of the least solution of 
a fix-point equation $\sigma^\sharp = F_{c_i}(\sigma^\sharp)$, $\sigma^\sharp \in \Sigma^\sharp$ where $\Sigma^\sharp$ 
is a domain of 
abstract properties, and $F_{c_i}$ is the abstract transformer for the cell, i.e., a composition of abstract 
statements transformers in the cell fix-point computation to solve the static analysis problem. We 
refer to~\cite{CC77} for a comprehensive background on abstract interpretation.

Within the abstract interpretation framework Several analysis can co-exist by constructing an \emph{independent 
product} of abstract domains. We denote executing several transformers in parallel for cell $c_i$ as $F_{c_i}^A$ 
where $A$ is a set of analyses. We refer the reader to~\cite{aiproduct} on literature on 
combining abstract domains with independent products.

\paragraph{Cell summary computation.} Apart from computing the abstract state, we compute 
\emph{pre-summaries}. Pre-summaries are intra-cell computed pre-conditions on a cell that are 
used to act as a pre-condition guard on if an abstract state should be propagated to that cell. We compute 
pre-summaries for each cell at notebook initialization time and during cell code changes.

In order to compute a pre-summary $pre_{c_i}$ for cell $c_i$ we construct use-def (U-D) structure
using standards data-flow techniques~\cite{dfa}. U-Ds provides a mappings between 
variable usages and their definitions. A variable is a defined if it is used as a right-hand-side 
expression in a assignment statement or if it is a function $st$. A variable is used if it is in a 
left-hand-side of an assignment statement or in a function $st$. Thus, given a cell $c$ we can 
define the following sets of variables that define definitions and usages.

\begin{align*}
def(c) =& \{ v \ | \ \forall \ st \in c \textit{ s.t. } v \textit{ is defined in } st \} \text{ and }\\
use(c) =& \{ v \ | \ \forall \ st \in c \textit{ s.t. } v \textit{ is used in } st \}
\end{align*}

The U-D structure is computed using a reaching definitions data-flow analysis~\cite{dfa} and 
provides a mapping $\textit{use-def}$ for all symbols $v \in V$ in the cell. If 
a $v \in use(c)$ has no definition, it is mapped to $\bot$. Using the U-D structure 
we compute the set of all unbound variables in a cell:
$$pre_{c_i} = \{ \ v \ | \ v \in use(c_i) \wedge \textit{use-def}(v) = \bot \} $$

Depending on the analysis we may want to expand the definition of $pre_{c_i}$. For example, 
for access violation we may want to ignore variables in cells where no access patterns occur and 
a variable is never used to change and propagate information e.g., simply printing data. 

\subsubsection{Inter Cell Analysis}
\label{ss:inter-technical}
\paragraph{State propagation.}
The inter-cell analysis computes a set of abstract states for the entire notebook up to a 
depth $K$ or to a global fixpoint solution (fixpoints on all active paths).  The abstract state from a source cell is propagated to other cells
if and only if there exists an edge that satisfies a cell propagation dependency. When the propagation occurs, 
an intra-cell analysis computation is performed that treats the incoming cell abstract state as the initial 
state.  

\begin{figure}
    \centering
    \includegraphics[width=0.45\textwidth]{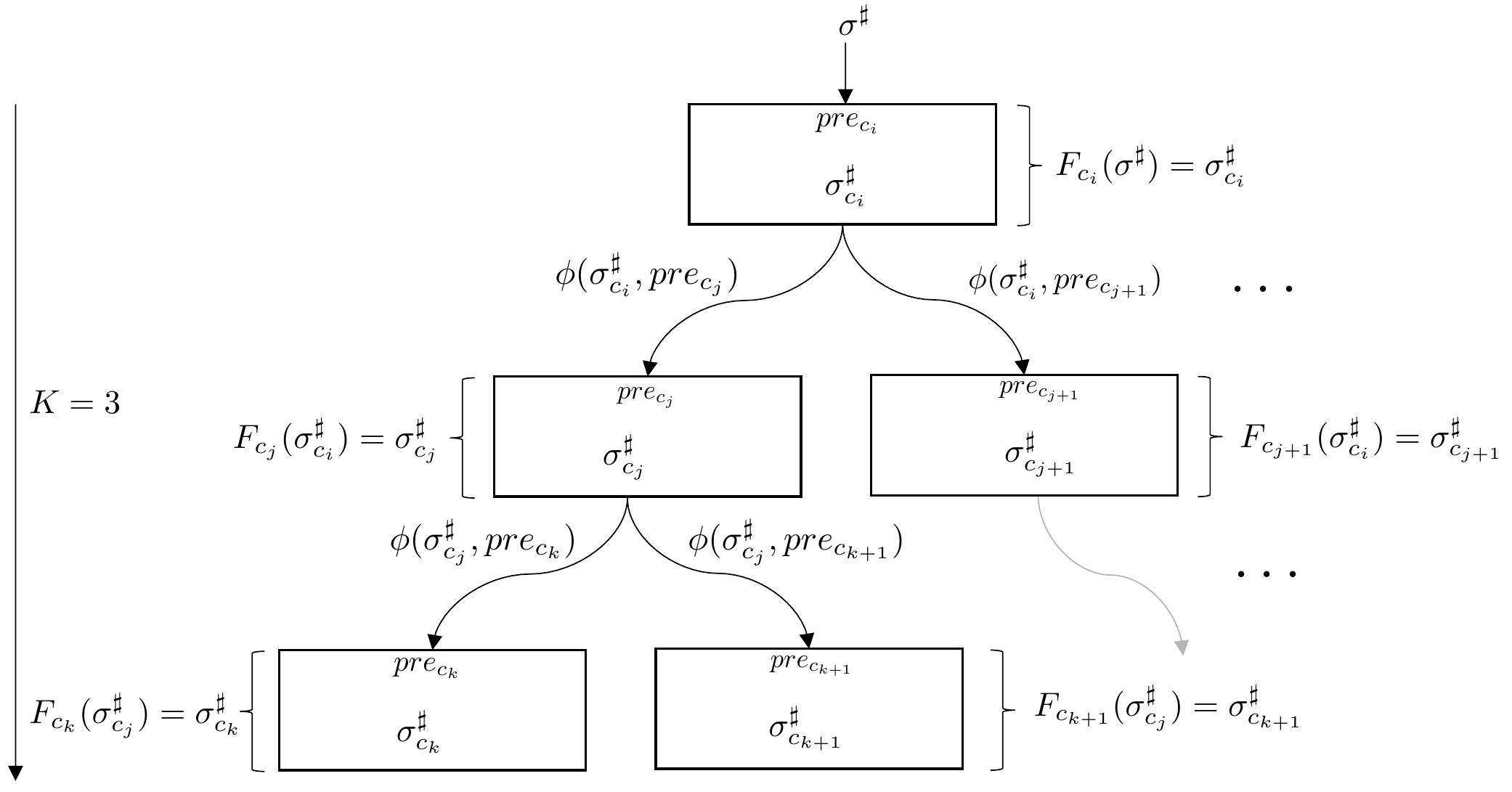}
    \caption{Inter Cell Analysis\label{fig:fw}}
\end{figure}

The inter-cell analysis propagation is depicted in Figure~\ref{fig:fw}. Here the 
what-if analysis is triggered by an event $e$ for a source cell 
$c_i$. A pre-defined value of $K \in \{1, \dots, \infty\}$ is defined where  $K=\infty$ 
means we compute until a fix-point, that determines the depth of the analysis. The dependency is defined by 
determining if the abstract state $\sigma^\prime_{c_i}$ of the cell $c_i$ can be combined with the pre-summary
$pre_{c_j}$ of another cell $c_j$ (which may be cell $c_i$ itself). If there is a dependency, the unbound 
variables in $c_j$ consume their values from $\sigma^\prime_{c_i}$. This propagation is continued until all paths 
terminate either by reaching the $K$ limit, by achieving a fixpoint or by $\phi$ not holding for all cells in the notebook.

We formalise cell dependencies in the form of a graph definition. Note, in our technique the graph is constructed 
lazily during abstract state propagation phase.

\begin{definition}[Cell Propagation Dependency Graph]
\label{def:celldepgraph}
We assume the sequence of cells form a directed dependency graph $\mathcal{G} = \langle N, D \rangle$
where $N$ is a finite set of cells, and $(c_i, R, c_j) \in \mathcal{R}$ defines an
arc from cell $c_i \in N$ to $c_j \in N$ iff $\phi(\sigma^\sharp_{c_i}, pre_{c_j})$. How 
$\phi(\sigma^\sharp_{c_i}, pre_{c_j})$ is defined is analysis specific. In Section~\ref{sec:analyses} 
we provide examples of how analyses can be defined to fit into the \nb framework.

\end{definition}


\paragraph{Fixpoint pruning.} The control-flow from inter-cell execution can result in cycles. For example, 
a cell can be re-executed as the dependency $\phi$ repeatedly holds in a given sequence. For analyses with $K=\infty$ 
this can cause non-termination. Even, in the case for bounded $K$ values, ignoring fixpoints is a source of inefficiency. Firstly, 
by fixpoint we mean that the state $\sigma^{\sharp}_{c_i} = \sigma^{\sharp'}_{c_i}$, that is, in an analysis path (note many paths exist 
as shown in Figure~\ref{fig:fw}), the state resulting from first execution of cell $c_i$ does not change after its next execution. There may or 
may not be execution of cells in between. Regardless, this implies that we can prune this execution path as other execution paths arising 
from the first execution of $c_i$ will subsume any further executions from the second execution of $c_i$. For analyses with high or 
infinite lattices, extrapolation operators are needed for force convergence. Therefore we detect when we re-visit a cell and 
terminate the exploration on that path if a fixpoint is detected after the re-execution. In the case of high or infinite abstract domains, we 
extrapolate the fixpoint using standard widening techniques~\cite{CC77} for the given abstract domain.

\begin{obs}[Fixpoint Subsumption]
Given a fixpoint on a cell execution path $p = \langle c_{i}, \dots, c_{i+k}\rangle$ such that $c_i = c_{i+k}$ and 
$\sigma^\sharp_{c_i} = \sigma^\sharp_{c_{i+k}}$, any cell $c_j$ not in $p$ executed after $c_i$ will produce the same abstract 
state as being executed after $c_{i+k}$.
\end{obs}
\begin{proof}
Since the $c_i$ and $c_{i+k}$ are the same cell and since we have a fixpoint 
i.e., $\sigma^\sharp_{c_i} = \sigma^\sharp_{c_{i+k}}$, we want to show that it follows 
the cells in between do not contribute to the computation. For a contradiction lets assume they do 
and a cell $c_{i+k-j}$ where $j < k$ executed after $c_i$ and 
before $c_{i+k}$. Due to the assumed monotonicity of the transformers,  
$\sigma^\sharp_{c_i} \neq \sigma^\sharp_{c_{i+k}}$ which would contradict our assumption of a fixpoint.
Therefore, no $c_{i+k-j}$ for all $j < k$ produced an non operation (nop) execution and can be pruned.
\end{proof}

\subsubsection{Algorithmic Implementation}

\begin{algorithm}[h]
    \caption{Event\_Handler($code, c_i, e, K$)}
    \label{alg:highlevel}
    \begin{algorithmic}[1]
    \Variables
    \State $\sigma^\sharp$ (global abstract state)
    \State $\textit{pre}$ (cell to pre-summary mapping)
    \State $\textit{cfg}$ (cell to cfg mapping)
    \EndVariables
    \State $A' = \mathcal{M}(e)$\label{line:getanalysis}
    \If{$A' = \emptyset \wedge e = \textit{execute}$}
      \State $\sigma^\sharp := \textit{Maintain}(code, c_i, \sigma^\sharp, A)$\label{line:maintain}
    \Else
      \State $\textit{report} := \textit{InterCell}(\sigma^\sharp, c_i, K, [], A', \eta)$\label{line:whatif}
      \State \Return $\textit{report}$
    \EndIf
    \end{algorithmic}
\end{algorithm}

Our technique is described algorithmically, starting with Algorithm~\ref{alg:highlevel} which receives an 
event and determines if the computation should proceed in maintenance mode or what-if analysis mode. Given 
an event $e$ occurs, we obtain the source cell code $code$, identifier $c_i$, event $e$ and global abstract 
state. At line~\ref{line:getanalysis}, we determine if there exist any analyses $A' \subseteq A$ that are attached 
to the event $e$. If not, we perform 
a maintenance in line~\ref{line:maintain} by calling 
$\textit{Maintain}(code, c_i, \sigma^\sharp, A)$ and updating the global. Otherwise, we proceed with a what-if analysis by calling 
$\textit{InterCell}(\sigma^\sharp, c_i, K, [], A', \eta)$ in line~\ref{line:whatif} and return the results of the 
analysis to the notebook. Here apart from the global state $\sigma^\sharp$, cell label $c_i$, and the analyses $A'$, 
the $K$ parameter is passed, along with a report which is initialized to an empty list $[]$ 
(to simplify the algorithm description we represent the report as simply a list), and a mapping $\eta: N \rightarrow \Sigma^\sharp$ which 
maps each cell label to its last computed abstract state. This is required to detect fixpoints in the analysis paths.

In Algorithm~\ref{alg:analyzeCell} we describe intra-cell analysis, namely cell maintenance. The function 
\textit{Maintain} first checks to see if a code change occurred. If so,  
the pre-summary $pre_{c_i}$ is re-built and an intra cell static analysis 
$F_{c_i}(\textit{cfg}_{c_i}, \sigma^\sharp)$ is performed to produce a new abstract state 
$\sigma^\sharp$. If the code has not changed, since the abstract state may have changed in the mean time, 
we perform an intra-cell analysis i.e., $F_{c_i}^A(\textit{cfg}_{c_i}, \sigma^\sharp)$ for all analyses in $A$. 
Note, that we cache CFGs, U-Ds, and pre-summaries to avoid unnecessary re-computation.

\begin{algorithm}[h]
    \caption{Maintain($code, c_i, \sigma^\sharp, A$) \label{alg:intracell}}
    \label{alg:analyzeCell}
    \begin{algorithmic}[1]
        \If{$code$ not changed }
          \State{$\sigma^{\sharp'} := F_{c_i}^{A}(\textit{cfg}_{c_i}, \sigma^\sharp)$}
          \State \Return $\sigma^{\sharp'}$
        \EndIf
        \State{\textit{ast := parse(code)}}
        \State{$\textit{cfg}_{c_i} := getCfg(ast)$}
        \State{\textit{ud := getUD(cfg)}}
        \State{$\textit{pre}_{c_i}$ := getPre(ud)}
        \State{$\sigma^{\sharp\prime} := F_{c_i}^{A}(\textit{cfg}_{c_i}, \sigma^\sharp)$}
        \State \Return $\sigma^{\sharp'}$
    \end{algorithmic}
\end{algorithm}

The \textit{InterCell} algorithm described in Algorithm~\ref{alg:whatif} performs the what-if analysis. Here 
analyses in $A'$ are executed on cells, starting with the source cell $c_i$ in lines~\ref{line:F} and \ref{line:check} 
and propagating the abstract state to 
cells that have a dependency i.e., satisfy $\phi(\sigma^{\sharp\prime}, pre_{c_j})$ as shown in 
lines~\ref{line:cond} and \ref{line:prop}. Note for each application of the transformer 
the initial state must be joined with the previous initial state i.e., $\sigma^\sharp \sqcup \eta_{c_i}$.
If $K = 0$ (line \ref{line:k}), meaning we have reached the 
required depth or a fixpoint is detected (line \ref{line:fp}) we terminate. The algorithm worst case complexity is 
$\mathcal{O}(n^K)$ in the number of cells $n$ and depth $K$.

\begin{algorithm}[h]
    \caption{InterCell($\sigma^\sharp$, $c_i$, $K$, $report$, $A'$, $\eta$)\label{alg:intercell}}
    \label{alg:whatif}
    \begin{algorithmic}[1]
         \If{$K = 0$}\label{line:k}
           \Return $\textit{report}$
         \EndIf

         \State $\sigma^{\sharp\prime} := F_{c_i}^{A'}(cfg_{c_i}, \sigma^\sharp \sqcup \eta_{c_i})$\label{line:F}
         \If {$\sigma^{\sharp\prime} = \eta_{c_i}$}\label{line:fp}
           \Return $[]$
         \EndIf
         \State $\eta_{c_i} := \sigma^{\sharp\prime}$
         \State $report' := Check(\sigma^{\sharp\prime}, A', report)$\label{line:check}
         \ForAll{$c_j \in N$}
            \If{$\phi(\sigma^{\sharp\prime}, pre_{c_i})$}\label{line:cond}
            \State report' := report' + \textit{InterCell}($\sigma^{\sharp\prime}$, $c_j$, $K-1$, report', A', $\eta$)\label{line:prop}
            \EndIf
         \EndFor
         \Return $\textit{report'}$
    \end{algorithmic}
\end{algorithm}

An optional operation, that we omit here, is to perform inter-cell widening. Widening is required for 
analyses with potential slow convergence such as interval analysis. Adding widening would require an 
extra condition in the code that checks if the abstract state increases on a given variable. If so, 
the value for that variable is added as the top element. A narrowing pass can be also performed to 
improve precision.

\subsubsection{Analysis Criteria and Contracts}
The \textit{Check} function in Algorithm~\ref{alg:intracell} and \ref{alg:intercell} 
checks the abstract state after a cell execution and depending on the criteria determine if 
a violation has occurred. For standard built in analyses (see Section~\ref{sec:analyses}) this is 
hard coded into \nb. However, for the available abstract domains, a user can define \emph{contracts} 
on lines of code, pre or post conditions on cells or on the global notebook. \nb exposes the set of available 
abstractions available, which can be seen as schema for which users can define queries in a logic based DSL 
that can assert expected behaviour.

The analysis provides a set of finite sets of objects from the AST and analysis results that the user can 
formulate as a error condition, attached to a notebook, cell or code line. Any languages that maps to first order logic (with 
finite domains) can be used. For example, Datalog or SQL are both candidates.

\section{Instantiated Analyses}
\label{sec:analyses}
In this section we give a brief outline of several instantiations of our analysis framework which we later 
evaluate in Section~\ref{sec:evaluation}.

\subsection{Use Case I:  Code Impact Analysis}
\label{sec:analysis-cia}
When a code change occurs users need to know what other code is affected or unaffected by that change. This has 
a number of usages including automating notebook execution, stale cell state detection, 
cell cleanup and simplification among others. 

\subsubsection{Abstract Semantics}
We define an abstract domain that maps a variable (including symbols, function names, etc. ) $v$ to a Boolean indicating 
which variable has changed or not. Practically, the abstract domain is implemented as a set of variables 
$\bar{v} \subseteq V$. If a variable 
is in the set it has changed, otherwise it hasn't. Thus the lattice is a standard powerset lattice $\wp(V)$ common in 
data flow analyses~\cite{dfa}. When a variable on the left-hand-side of a statement has changed, we insert the 
right-hand-side in the set. Below we state the propagation semantics for selected statements. In the implementation, 
more variations are covered.

\begin{enumerate}
\item Assignment:
\begin{align*}
  \lambda \sigma^\sharp.\llbracket \bar{y} = f(\bar{x}) \rrbracket  = \braces{
    \begin{array}{l}
      \sigma^\sharp \cup \{y\} \text{ iff }  f \in \sigma^\sharp \vee \\ 
      \exists x \in \bar{x} \text{ s.t. } x \in \sigma^\sharp \\
    \end{array}
  }
\end{align*}

\item Functions:
\begin{align*}
  \lambda \sigma^\sharp.\llbracket f(\bar{x}) \{ \bar{y} \} \rrbracket  = \braces{
    \begin{array}{l}
      \sigma^\sharp \cup \{ f \} \text{ iff }  \\
      \exists x \in \bar{x} \text{ s.t. } x \in \sigma^\sharp \vee \\
      \exists y \in \bar{y} \text{ s.t. } y \in \sigma^\sharp \\
    \end{array}
  }
\end{align*}

\end{enumerate}

Similarly, joins and meets that arise from control flow are handled by the join operations of the abstract domain, i.e., 
set union and disjunction. 

The $\phi$ operation in this case holds if the intersection of 
the variable in $\sigma^\sharp$ and $pre_{c_j}$ are not null, i.e., there is at least one dependent variable in $c_j$ 
to propagate to. More formally, we define $\phi$ as follows:

$$\phi(\sigma^\sharp_{c_i}, pre_{c_j})  = \{v :  v \in \sigma^\sharp_{c_i} \} \cap pre_{c_j} \neq \emptyset$$

This analysis has many use cases including automating cell executions for a given change by following the 
dependencies; stale cell execution where cells that have intermediate impacted cells between them and the source cell,
can potentially be stale. Typically, we execute this analysis with a bounded e.g., $K=3$ for stale cell analysis but in 
principle can be unbounded. Both options result is similar overheads due to fixpoint subsumption. Other variations include 
fresh cell analysis that detects cells that cannot cause staleness when changed. As well as, 
isolated cell analysis that detects cells without dependencies on other cells and thus can be potentially cleaned. These cells 
are typically found during experimentation phases of development and need to be identified when the notebook program is 
converted to a production script. This analysis is only performed on $K=1$ due to its nature.






\subsubsection{Analysis Example}
Consider the example in Figure~\ref{fig:motivating}. Here we can see that the execution of cell 1 followed by cell 4 will create staleness. This 
is because cell 2 is fresh and is the intermediate cell between cell 1 and cell 4 dependencies. Suppose we change the file in cell 1 then the variable 
$d$ is in our abstract domain. As before we propagate this to cell 2 and hence $x$ is also in our abstract domain. When we further propagate to 
cell 4 (i.e., $K = 2$) we can report that all the right-hand-side variables are stale if the cell execution sequence 1, 4 is performed.

\subsection{Use Case II:  ML Data Leakage}
\label{sec:analysis-dl}
Data Leakage~\cite{dataleak} is a common bug specific to data science. In machine leaning applications 
a models typically require normalization of the input data, especially neural networks. Commonly, data is normalized by 
performing a division of the existing data by its average or maximum. Likewise, data is typically split into 
training and test subsets. If the normalization is performed using 
the \emph{overall} data set, then information from the test set will now be influencing the training subset. For this reason, 
any normalization should be applied individually on the test and training subsets.

While the Data leakage commonly occurs in data science scripts it is even further exacerbated by the 
execution semantics of notebooks. To this end, we have implemented a light-weight analysis to detect 
potential data leakages in notebooks. Our abstraction tracks which variable points to which data source and if the 
variable has been used to train or test a model. When an 
operation is performed on data that can introduce a leak, e.g., normalization, extrapolation etc. we reset the 
data source propagation. Otherwise we propagate the source dependencies of left hand side variables to 
right hand side variables. When a variable is an argument to a function that is marked as training or testing 
a model, the variable is marked as such. We assert that no two variables that are marked as train and test, respectively
point to the same data source.

\subsubsection{Abstract Semantics}
We define an abstract domain that maps a variable $v$ to a data leakage abstract domain $\langle L,\sqsubset \rangle$ where 
$L = \wp(v) \times \wp(\{tr, ts\})$ such that $L$ is partially ordered ($\sqsubset$) point wise by the subset 
relation $\subset$ with meet $\sqcup$ and join $\sqcap$ are similarly defined using point wise set union 
$\cup$ and intersection $\cap$, respectively. Thus, an element in the abstract domain is a variable that maps 
to a set of variables and an indicator if it has been an argument in a train or test function (or both). We 
differentiate between which tuple element in the product domain is access by $\sigma^{\sharp1}$ (first tuple element) 
and $\sigma^{\sharp2}$ (second tuple element). We define a simplified abstract semantics for three categories of operations 
below. Note, we do not exhaustively cover all cases below for readability, and only highlight the most important cases.
\begin{enumerate}
\item reset:
\begin{align*}
  \lambda \sigma^\sharp.\llbracket \bar{y} = f(\bar{x}) \rrbracket  = \braces{
    \begin{array}{l}
      \forall y \in \bar{y}.\sigma^\sharp[y \mapsto (\bar{x}, \emptyset)] \\
      \text{ iff } f \in KB_{\text{reset}}
    \end{array}
  }
\end{align*}

\item propagate:
\begin{align*}
  \lambda \sigma^\sharp.\llbracket \bar{y} = f(\bar{x}) \rrbracket  = \braces{
    \begin{array}{l}
      \forall y \in \bar{y}.\sigma^\sharp[ y \mapsto \sigma^\sharp(y) \sqcup \bigsqcup_{x \in \bar{x}} \sigma^\sharp(x)] \\
      \text{ iff } f \notin KB_{\text{reset}}, KB_{\text{test}}, KB_{\text{train}}
    \end{array}
  }
\end{align*}

\item sinks:
\begin{align*}
  \lambda \sigma^\sharp.\llbracket f(\bar{x}) \rrbracket  = \braces{
    \begin{array}{l}
      \forall x \in \bar{x}.\sigma^\sharp[x \mapsto (\sigma^{\sharp1}(x),  \sigma^{\sharp2}(x) \sqcup \{tr\})]
      \\ \text{ iff } f \in KB_{\text{train}} \\
      \forall x \in \bar{x}.\sigma^\sharp[x \mapsto (\sigma^{\sharp1}(x),  \sigma^{\sharp2}(x) \sqcup \{ts\})]
      \\ \text{ iff } f \in KB_{\text{test}} \\
    \end{array}
  }
\end{align*}
\end{enumerate}

The reset operations (1), occur when a function f is marked as a reset function in the knowledge base KB. This is 
library specific and rely on the data scientists to specify which functions may cause leakage. Here we map 
each right hand side variable $y$ to the variables that are arguments to the reset function while resetting 
its markings of test or train usage to $\emptyset$.

In the case (2), we simply propagate information by joining all abstract states of right-hand-side variables to 
the left-hand-side variables existing abstract state.

Finally in the case of (3), we mark each argument with a \{tr\} element if it is an argument to a train function, and \{ts\} 
if it is an argument to a test function. If a variable marked as \{tr\} points to the same source as a function marked as 
\{ts\} we issue a warning.

To enable inter-cell propagation we define the rule:

\begin{align*}
  \phi(\sigma^\sharp_{c_i}, pre_{c_j}) =& pre_{c_j} \subseteq  \{v :  (v \mapsto x) \in \sigma^\sharp_{c_i} \wedge x \neq \bot \} \\  
                                           & \wedge pre_{c_j} \neq \emptyset
\end{align*}


Here we make sure the successor cells pre is not empty and is a subset of the incoming cells abstract state variables 
that are reachable (not mapped to $\bot = (\emptyset, \emptyset)$). This analysis can be performed on a variety of 
sizes of $K$, however in practice we rely on our fixpoint subsumption and keep $K$ unbounded. 

\subsubsection{Analysis Example}
Consider the example in Figure~\ref{fig:motivating}. Recall,
in example, a potential data leakage is shown. Depending on the execution 
order of the cells i.e., if \textit{cell 2} is executed before \textit{cell 4} and $5$. We now describe how 
our analysis will detect this violation. Lets assume a what-if analysis is triggered 
for for the event of executing \textit{cell 1}. In other words, we ask \emph{what can happen if in 
future executions if cell 1 is executed?} We first compute an abstract state for \textit{cell 1} which is 
$\sigma^\sharp_{c_1} = d \mapsto (\{ data.csv \}, \emptyset)$. Now using abstract state and preconditions 
of other cells we asses the value of 
$\phi(\sigma^\sharp_{c_1}, pre_{c_j})$ for all cells $c_j$ in the notebook. We find 
$\phi$ holds for \textit{cell 2}. Next, we compute the abstract state for 
\textit{cell 2} with the abstract state of \textit{cell 1} as the initial state, obtaining
$\sigma^\sharp_{c_2} = d \mapsto (\{ data.csv \}, \emptyset), x \mapsto (\{d \}, \emptyset)$ as we apply rule (2)
as $\textit{fit\_transform} \in \textit{KB}_{reset}$. We evaluate $\phi(\sigma^\sharp_{c_2}, pre_{c_j})$ 
for all cells $c_j$ in the notebook and find that \textit{cell 4} holds. Here all split variables 
map to $d$ applying rule (1). Again we find that we can only propagate to \textit{cell 5}. 
Here we have a function in 
$\textit{fit} \in \textit{KB}_{train}$ so we have that $\textit{x\_train} \mapsto (d, \{tr\})$ and 
$\textit{y\_train} \mapsto (d, \{tr\})$. Next, we have a function $\textit{predict} \in KB_{test}$ 
and so we have $\textit{x\_test} \mapsto (d, \{ts\})$. Since we have $2$ or more variables that point to 
$d$ and have $\{tr\}$ and $\{ts\}$ we flag \textit{cell 5} asserting that the 
following conditions is violated: \emph{no arguments of train and test functions can point to the same data}. With 
this analysis condition, \nb warns the 
user that the execution sequence of cells executions $\langle 1, 2, 4, 5 \rangle$ will result in a 
data leakage in \textit{cell 5}.

\section{Integration and Applications}
We have implemented our technique as a notebook server extension. The client and server communicate through 
communication channels, where the client advises the server of events and sends code to be executed.  The server 
performs the static analysis. When the analysis is complete, 
we send information back to the client that contains affected cell ids, cell sequences, line numbers, messages, telemetry etc. to warn the user 
by altering the client graphical user interface.
Our implementation currently targets the python language. It parses the code into an AST from which it constructs a 
CFG and U-D chains. These low level code representations are used to perform the analyses 
implemented in our framework. The user can manually trigger the what-if analysis and pre-select which built-in analyses 
are turned on for what event. We warn the user of potential violations through use of cell highlighting and messages. The user 
interface varies with the clients used.

The standard use case is to use \nb as a development time what-if advisor that detects potential bugs before they 
occur. However \nb can be used for notebook cleaning, where erroneous, idle and isolated cells are flagged for removal;
notebook reproducibility where safe execution paths that lead to acceptable results are extracted from the notebook; 
auditing where security violations for e.g., GDPR compliance~\cite{gdpr}; education where the tool can help data scientists 
write quality code.

\section{Evaluation}
\label{sec:evaluation}
In this section, we evaluate \nb on a set of real world notebook benchmarks and on the instantiated analyses in 
Section~\ref{sec:analyses}. The evaluation aims to \textbf{(1) validate the adequacy of the \nb's performance for our 
notebook use cases} and to \textbf{(2) investigate performance bottlenecks and configuration tuning required to maximize in the performance of \nb}.  As a universal 
service level agreement (SLA) model for user experiences, i.e., the acceptable delays the analyses may exhibit in the notebook environment we refer to the RAIL model~\cite{rail}.

\subsection{Experimental Setup}
All experiments were performed in an Intel(R) Xeon W-2265 CPU @ 3.50 GHz with 64 GB RAM running a 64 bit Windows 10 operating 
system. Python 3.8.8 was used to execute \nb. We evaluate the execution-time of our static analyses running within 
\nb on the full set of Kaggle notebooks.

\subsubsection{Static Analyses.}
We perform the static analyses instantiated in our framework below. For each analysis type 
we perform an analysis over all notebooks, where each cell is a source.

\begin{itemize}[leftmargin=*]
\item Code Impact Analysis (CIA): 
the following variations of the analysis discussed in Section~\ref{sec:analysis-cia}:
\begin{itemize}
\item Isolated Cell Analysis (IsC) 
\item Fresh Cell Analysis (FrC) 
\item Stale Cell Analysis (StC) 
\end{itemize}
\item Data Leakage Analysis (DlK): the analysis discussed in Section~\ref{sec:analysis-dl} 
\end{itemize}

\subsubsection{Benchmark Characteristics.} We use a benchmark suite consisting for $2211$ executable real world 
notebooks from the Kaggle competition\cite{kaggle} that has previously used to evaluate data science static 
analyzers~\cite{vamsa}. The benchmark characteristics are summarized in Table~\ref{tbl:bm}.

We see on average, the notebooks in the benchmark suite have $24$ cells, where each cell on average has 
$9$ lines of code. In addition, on average branching instructions appear in $33$\% of cells. Each notebook has on 
average $3$ functions and $0.1$ classes defined. We note that these characteristics of low amount of branching, 
functions and classes is typically advantageous for static analysis precision. We found that every second notebook 
had a cell that could not be parsed and analyzed due to a syntax errors in it. Overall, this affected 4\% of cells in the 
benchmarks. $2.1$ of variables were unbound, from an average of $8.2$ variables per cell. 

\begin{table}[]
\tiny
\begin{tabular}{|l|l|l|l|l|}
Characteristic                    & Mean  & SD  & Max  & Min \\ \hline
Cells (per-notebook)              & 23.58 & 20.21  & 182  & 1 \\
Lines of code (per-cell)          & 9.12 & 13.55  & 257  & 1 \\
Branching instructions (per-cell) & 0.43  & 2.49 & 76 & 0 \\
Functions (per-notebook)          & 3.33 & 7.11 & 72 & 0 \\
Classes (per-notebook)            & 0.14  &0.64  & 11 & 0 \\
Non-parsing cells (per-notebook)  & 0.5 & 0.98 & 20 & 0 \\
Variables (per-cell)              & 8.2 & 2.3 & 552 & 0 \\
Unbound variables (per-cell)    & 2.1 & 1.06 & 12 & 0 \\ \hline
\end{tabular}
\caption{Kaggle Notebook Benchmark Characteristics\label{tbl:bm}}
\end{table}

\subsection{Code Change Analysis} 
We evaluate the performance of the code impact analyses for notebook usage. Recall the CIA analyses 
reason about change propagation in notebook cells. Also recall that many of these analysis 
by definition have a \textbf{fixed $K$}. For evaluation purposed 
we also evaluate stale cell analysis for various values of $K$ including $K=\infty$, since it can be run with such 
a parameter value.  For each evaluation of CIA analysis, we perform a random change in an source cell and compute 
the analysis result. We do this for all cells in a notebook, and for all notebooks in the benchmark suite. We omit notebook cells 
that do not parse. 

\subsubsection{Isolated Cells.} 
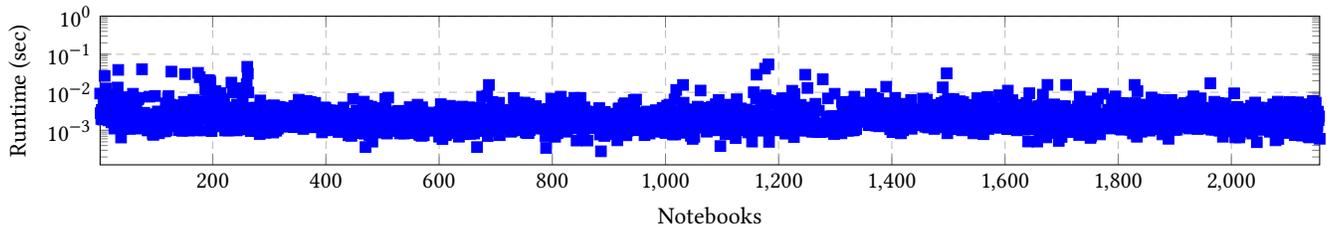
\begin{figure*}
\begin{tikzpicture}
\begin{axis}[
width=\textwidth,
height=0.20\textwidth,
ymode=log,
xlabel={Notebooks},
ylabel={Runtime (sec)},
xmin=1, xmax=2156, 
ymin=0, ymax=1,
xmajorgrids=true,
ymajorgrids=true,
grid style=dashed,
]

\addplot[only marks, blue, mark=square*] table [x=n, y=is, col sep=tab]{data/boundedtimes.csv};
\end{axis}
\end{tikzpicture}
\caption{Isolated Cell Analysis Avg. Times\label{perf:iso}}
\end{figure*}

In Figure~\ref{perf:iso} the average analysis executions per-notebook are shown. Overall, 
the results how that the average isolation analysis on a notebook takes 2.499 milliseconds.  The 
average analysis time, including the maximum outliers, are \textbf{well under the threshold for humans to 
notice a delay} and does not degrade the user experience. We were able to detect that 14\% of cells were in 
fact isolated and candidates for notebook cleaning or simplification. \textbf{This highlights the need for 
automated cleaning/simplification tooling for notebooks} and the ability of \nb to detect such cells. Our 
manual observation of the results could not detect any false positives.


\subsubsection{Fresh Cell Analysis}
\begin{figure*}
\begin{tikzpicture}
\begin{axis}[
width=\textwidth,
height=0.20\textwidth,
ymode=log,
xlabel={Notebooks},
ylabel={Runtime (sec)},
xmin=1, xmax=2156, 
ymin=0, ymax=1,
xmajorgrids=true,
ymajorgrids=true,
grid style=dashed,
]
\addplot[only marks, blue, mark=square*] table [x=n, y=fr, col sep=tab]{data/boundedtimes.csv};
\end{axis}
\end{tikzpicture}
\caption{Fresh Cell Analysis Avg. Analysis Times\label{perf:fresh}}
\end{figure*}
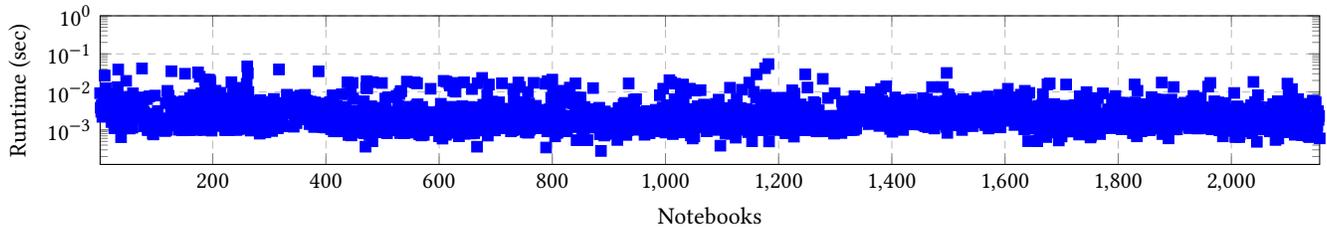

In Figure~\ref{perf:fresh} the average fresh cell analysis executions are shown. The results show that the 
average fresh cell analysis on a notebook takes 2.98 milliseconds. Similarity, the analysis time for average 
and maximum outliers is \textbf{well under the threshold for users to notice a delay} and does not degrade the 
user experience. The analysis found that 74\% of cells are fresh and no false positives were observed by our 
manual inspection.

\subsubsection{Stale Cell Analysis}
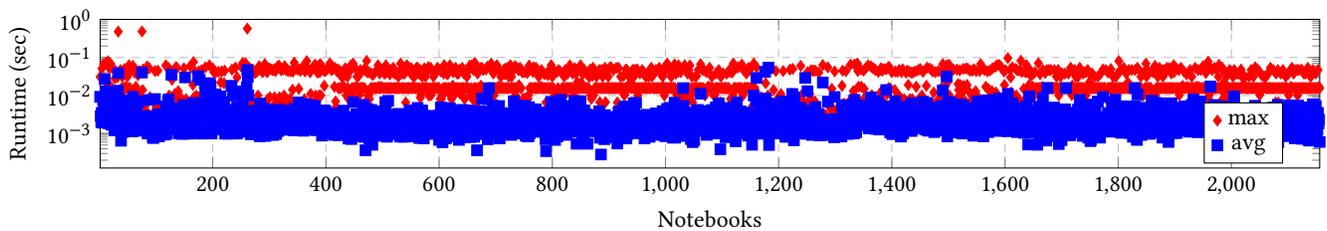
\begin{figure*}
\begin{tikzpicture}
\begin{axis}[
width=\textwidth,
height=0.20\textwidth,
legend pos=south east,
ymode=log,
xlabel={Notebooks},
ylabel={Runtime (sec)},
xmin=1, xmax=2156, 
ymin=0, ymax=1,
xmajorgrids=true,
ymajorgrids=true,
grid style=dashed,
]
\addplot[only marks, red, mark=diamond*] table [x=n, y=ms, col sep=tab]{data/boundedtimes.csv};
\addlegendentry{max}

\addplot[only marks, blue, mark=square*] table [x=n, y=as, col sep=tab]{data/boundedtimes.csv};
\addlegendentry{avg}
\end{axis}
\end{tikzpicture}
\caption{Stale Cell Analysis Avg. and Max Analysis Times\label{perf:stale}}
\end{figure*}

In Figure~\ref{perf:stale} the average and maximum stale analysis executions are shown. We show maximum results 
because we ran the experiments for $K=3$ and thus a propagation occurs. The results show that the average stale 
cell analysis on a notebook takes 2.49 milliseconds. The average maximum analysis per notebook take 
22.1 milliseconds, with a global maximum of 472.99 milliseconds. The analysis time for average and maximum 
cases are \textbf{well under the threshold for users to notice a delay} and does not degrade the user 
experience. While the global maximum analysis time is above the unnoticeable threshold, \textbf{it is still well under the threshold of any degradation of the user experience}. The analysis 
found that $24$\% of cells can cause staleness from a random change, highlighting the \textbf{significant possibility of such bugs occurring} and 
utility of this analysis. Our manual inspection found no false positives.

\subsection{Data Leakage Cell Analysis}
We evaluate the performance and utility of the data leakage analysis. This analysis is run on $K=\infty$, however we 
investigate its performance on various fixed $K$ values. This analysis is also diverse in its 
use cases. On one hand it can be used during development time to warn users of potential bugs and on the other hand it 
can be used in a batch mode for semantic reproducibility of notebooks~\cite{rep}. We evaluate if the performance 
of this analysis is compatible with these use cases.

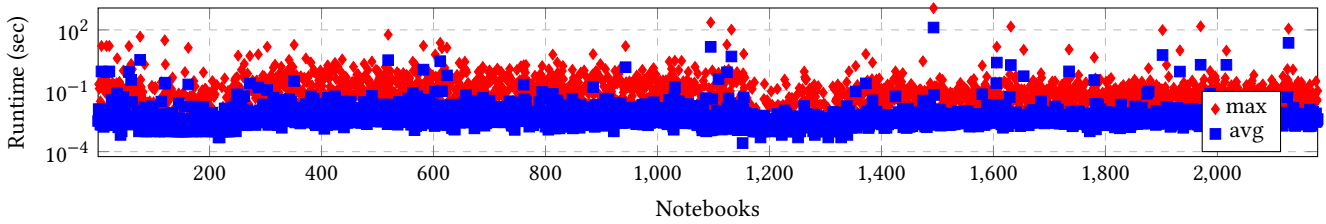
\begin{figure*}
\begin{tikzpicture}
\begin{axis}[
width=\textwidth,
height=0.20\textwidth,
legend pos=south east,
ymode=log,
xlabel={Notebooks},
ylabel={Runtime (sec)},
xmin=1, xmax=2179, 
ymin=0, ymax=1200,
xmajorgrids=true,
ymajorgrids=true,
grid style=dashed,
]
\addplot[only marks, red, mark=diamond*] table [x=n, y=m, col sep=tab]{data/full.csv};
\addlegendentry{max}

\addplot[only marks, blue, mark=square*] table [x=n, y=a, col sep=tab]{data/full.csv};
\addlegendentry{avg}
\end{axis}
\end{tikzpicture}
\caption{Data Leakage Analysis Avg. and Max. Analysis Times\label{perf:dl}}
\end{figure*}

In Figure~\ref{perf:dl} the average and maximum data leakage analysis for $K=\infty$ executions are shown. The results how that 
the average data leakage analysis on a notebook takes 
41.45 milliseconds. The average maximum analysis per notebook take 880.9 milliseconds, with a global maximum of 
233 seconds. \textbf{The analysis time for average case is well under the threshold for users to 
notice any delay} and does not degrade the user experience. \textbf{The average maximum recorded analysis time is above the immediate fell threshold, 
but below the threshold for the task feeling out of flow} (1000ms). The global maximum does cause a considerable delay and 
user degradation. Below we investigate the cause and remedies for such cases but note they are not common and
only 4\% of all analyses execute for more than 1000ms and only 1\% for more than 5000ms.

\begin{figure}
\begin{subfigure}[b]{0.48\linewidth}
\begin{tikzpicture}
\tikzstyle{every node}=[font=\small]
\begin{axis}[
width=\textwidth,
height=0.8\textwidth,
legend style={font=\fontsize{4}{5}\selectfont, at={(1,0.6)}},
xlabel={K},
ylabel={Avg. Runtime (sec)},
xmin=0, xmax=4, 
ymin=0, ymax=0.8,
xtick={0,1,2,3,4},
xticklabels={$1$, $2$, $4$, $8$, $\infty$},
xmajorgrids=true,
ymajorgrids=true,
grid style=dashed,
]
\addplot[red, mark=diamond*] table [x=K, y=max, col sep=tab]{data/Kspeedups.csv};
\addlegendentry{max}
\addplot[blue, mark=square*] table [x=K, y=avg, col sep=tab]{data/Kspeedups.csv};
\addlegendentry{avg}
\end{axis}
\end{tikzpicture}
\caption{Avg. Data Leakage Runtime for K\label{k:dlrt}}
\end{subfigure}\hfill
\begin{subfigure}[b]{0.48\linewidth}
\begin{tikzpicture}
\tikzstyle{every node}=[font=\small]
\begin{axis}[
width=\textwidth,
height=0.8\textwidth,
legend style={font=\fontsize{4}{5}\selectfont, at={(1,0.6)}},
xlabel={K},
ylabel={Avg. Runtime (sec)},
xmin=0, xmax=4, 
ymin=0, ymax=0.5,
xtick={0,1,2,3,4},
xticklabels={$1$, $2$, $4$, $8$, $\infty$},
xmajorgrids=true,
ymajorgrids=true,
grid style=dashed,
]
\addplot[red, mark=diamond*] table [x=K, y=max, col sep=tab]{data/KspeedupsCIA.csv};
\addlegendentry{max}
\addplot[blue, mark=square*] table [x=K, y=avg, col sep=tab]{data/KspeedupsCIA.csv};
\addlegendentry{avg}
\end{axis}
\end{tikzpicture}
\caption{Avg. CIA Runtime for K\label{k:ciart}}
\end{subfigure}
\begin{subfigure}[b]{0.48\linewidth}
\begin{tikzpicture}
\tikzstyle{every node}=[font=\small]
\begin{axis}[
width=\textwidth,
height=0.8\textwidth,
legend pos=south west,
legend style={font=\fontsize{4}{5}\selectfont},
xlabel={K},
ylabel={Avg. Subsumptions},
xmin=0, xmax=4, 
ymin=0, ymax=60,
xtick={0,1,2,3,4},
xticklabels={$1$, $2$, $4$, $8$, $\infty$},
xmajorgrids=true,
ymajorgrids=true,
grid style=dashed,
]
\addplot[blue, mark=square*] table [x=K, y=avg, col sep=tab]{data/Ksubsumption.csv};
\end{axis}
\end{tikzpicture}
\caption{Avg. Subsumptions for K\label{k:sub}}
\end{subfigure}
\begin{subfigure}[b]{0.48\linewidth}
\begin{tikzpicture}
\tikzstyle{every node}=[font=\small]
\begin{axis}[
width=\textwidth,
height=0.8\textwidth,
legend pos=south west,
legend style={font=\fontsize{4}{5}\selectfont},
xlabel={K},
ylabel={Avg. Nodes},
xmin=0, xmax=4, 
ymin=0, ymax=6000,
xtick={0,1,2,3,4},
xticklabels={$1$, $2$, $4$, $8$, $\infty$},
xmajorgrids=true,
ymajorgrids=true,
grid style=dashed,
]
\addplot[blue, mark=square*] table [x=K, y=avg, col sep=tab]{data/Kprop.csv};
\end{axis}
\end{tikzpicture}
\caption{Avg. Number Nodes for K\label{k:nodes}}
\end{subfigure}\hfill
\begin{subfigure}[b]{0.48\linewidth}
\begin{tikzpicture}
\tikzstyle{every node}=[font=\small]
\begin{axis}[
width=\textwidth,
height=0.8\textwidth,
legend pos=south west,
legend style={font=\fontsize{4}{5}\selectfont},
xlabel={K},
ylabel={Avg. Max. Depth},
xmin=0, xmax=4, 
ymin=0, ymax=6,
xtick={0,1,2,3,4},
xticklabels={$1$, $2$, $4$, $8$, $\infty$},
xmajorgrids=true,
ymajorgrids=true,
grid style=dashed,
]
\addplot[blue, mark=square*] table [x=K, y=avg, col sep=tab]{data/Kdepth.csv};
\end{axis}
\end{tikzpicture}
\caption{Avg. Max. Depth for K\label{k:depth}}
\end{subfigure}\hfill
\begin{subfigure}[b]{0.48\linewidth}
\begin{tikzpicture}
\tikzstyle{every node}=[font=\small]
\begin{axis}[
width=\textwidth,
height=0.8\textwidth,
legend pos=south west,
legend style={font=\fontsize{4}{5}\selectfont},
xlabel={K},
ylabel={Avg. Reach K Limit},
xmin=0, xmax=4, 
ymin=0, ymax=50,
xtick={0,1,2,3,4},
xticklabels={$1$, $2$, $4$, $8$, $\infty$},
xmajorgrids=true,
ymajorgrids=true,
grid style=dashed,
]
\addplot[blue, mark=square*] table [x=K, y=avg, col sep=tab]{data/KLimit.csv};
\end{axis}
\end{tikzpicture}
\caption{Avg. Reached Limit for K\label{k:lim}}
\end{subfigure}

\caption{Avg. Performance for Benchmarks for K Values\label{perf:k}}
\end{figure}
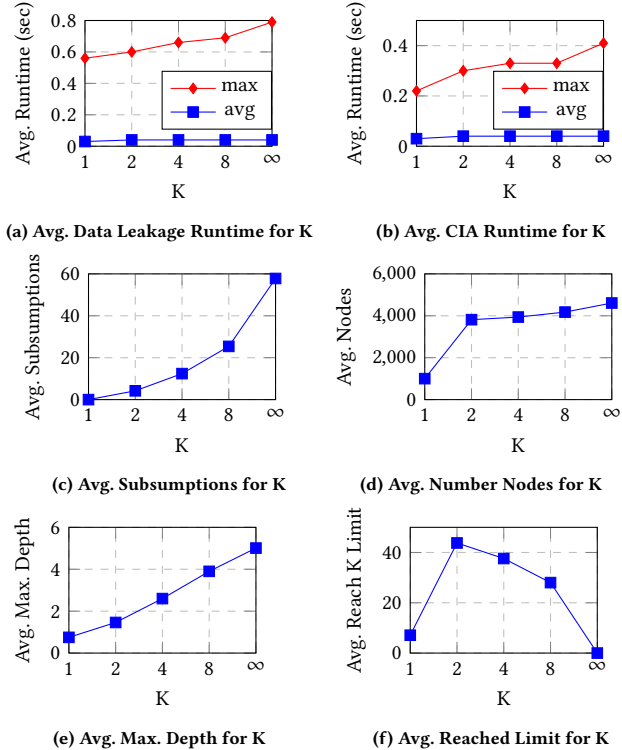

\begin{figure}
\begin{tikzpicture}
\begin{axis}[
width=\linewidth,
height=0.6\linewidth,
xlabel={Notebooks},
ylabel={Avg. Propagation Rate},
xmin=1, xmax=2179, 
ymin=0, ymax=0.9,
xmajorgrids=true,
ymajorgrids=true,
grid style=dashed,
]
\addplot[only marks, red, mark=diamond*] table [x=n, y=m, col sep=tab]{data/branch.csv};
\addlegendentry{max}

\addplot[only marks, blue, mark=square*] table [x=n, y=a, col sep=tab]{data/branch.csv};
\addlegendentry{avg}
\end{axis}
\end{tikzpicture}
\caption{Data Leakage Analysis Avg. and Max. Propagation Rates\label{perf:branch}}
\end{figure}
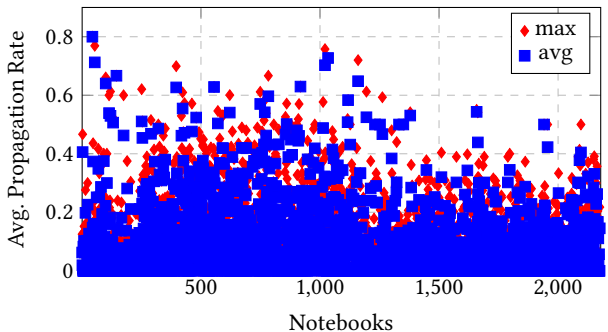

\subsubsection{Varying $K$ Bounds}
To demonstrate the affect of reducing $K$ in Figures~\ref{k:dlrt} and \ref{k:ciart} we show the effect of vary 
the $K$ from $\infty$ to bounded values of $K = 8, 4, 2$ and $1$. We note that the speedup of the analysis 
with smaller $K$ values \textbf{is not significant enough on average to result in major 
performance improvements} overall for either analysis. A major reason for this as shown in 
Figure~\ref{k:sub} is that our fixpoint subsumption optimization 
increases for large values of $K$ resulting in modest average maximum depths as shown in Figure~\ref{k:depth}. This is 
further corroborated by the Figures~\ref{k:nodes} and \ref{k:lim}. 

\subsubsection{Propagation Rate}
Another explanation for the achieved performance is the rate at which propagation occurs to other cells. Given a 
worst case scenario this propagation rate could be large and result in an execution tree that is very 
wide and deep. In Figure~\ref{perf:branch} we plot the average and maximum propagation 
rates for each notebook (where each cell was a source cell). Here we find that $\phi$ holds on average $9$\% of the time 
and the maximum propagation rate for each notebook is 
on average $13$\% of the time. This figure coincides with our experience \textbf{that a typical notebook has a several 
alternate executions but not to the extent that all cells depend on each other}.

\subsubsection{Bottleneck Sources}
Next we investigate the $4$\% of notebooks that take more than 1000ms to execute to determine where the bottleneck 
is occurring and what can be done to bring them under 1000ms. For the 103 notebooks with a maximum analysis 
execution of larger than 1000ms, we reduce $K = 4$. Here we were able to bring $33$\% (34 notebooks) of the 
notebook executions to under the 1000ms. \textbf{Thus, we are able to compute 98.7\% of notebooks in under 
1000ms}. We note, that this reduction exclusively 
occurred for notebooks with maximum execution times 
$<10$ seconds. Of the $4$ cases $>100s$ we found that they in fact had \textbf{no significant speedups by reducing 
the $K$ value}. We have investigated these notebook in question and narrowed down the poor performance to a 
single cells with very large number and depth of branching and large number of lines of code. For example, 
one of the notebooks had a cell with 257 LOC and $71$\% of these we 
branching statements often deeply nested. We argue that such code is not characteristic of regular notebooks 
and would cause a challenge to most static analyzers for the time frames we target. To 
mitigated such cases we would need to improve our rather standard intra-cell analysis fixpoint iteration 
techniques and employ techniques such as~\cite{souffle} which is left for future work.

\section{Related Work}
Compared to static analysis tools for data science scripts~\cite{sads}, \nb targets 
notebooks and incorporates their semantics in the analysis. We do this by deciding on-the-fly which cell 
should be analysed next depending on our cell propagation dependency graph. The use of static provenance is related 
to our technique in the sense that provenance information is propagated forward and can be modeled as an abstract 
domain~\cite{vamsa}. Unlike \nb this work targets scripts and is limited to loop free programs. The lineage 
technique in~\cite{nbsafety} targets notebooks. Using this static 
lineage information computed by data flow analysis and runtime time-stamps the technique is able to perform stale state 
detections. This technique relies on both runtime information and compile time (liveness data-flow analysis). \nb is can do 
such analyses statically and does not requiring information at cell execution. Moreover this technique is limited to 
a single analysis where \nb is a general static analysis framework and can incorporate various abstract domains.
The technique in~\cite{rep} attempts to reconstruct executions in notebooks by finding dependencies using syntactic 
means on the AST. \nb builds dependencies wrt. the abstract domain and performs a semantic analysis. Regarding our ML 
data leakage analysis, we have not found any static analysis method for detecting data leaks in data science code.


\section{Conclusion}
In this paper we have described \textsc{NBLyzer}, a static analysis framework that  
that takes into account the execution semantics of notebook environments. We have instantiated 
several analyses in our framework, with applications including notebook debugging, notebook verification, notebook cleaning, and 
notebook reproducibility. As far as we are aware, we are the first to suggest a general abstract interpretation-based 
static analysis framework specific for notebook semantics.

\section*{Acknowledgement}
We thank our colleagues at Gray Systems Lab (GSL), for making available the 
benchmarks suite used in~\cite{vamsa}. We thank our colleagues from 
Azure Data Labs for their feedback.

\bibliographystyle{ACM-Reference-Format}
\bibliography{refs}

\end{document}